\documentclass[12pt]{article}
\usepackage{amssymb}
\newcommand{\beq}{\begin{eqnarray}}
\newcommand{\eeq}{\end{eqnarray}}

\newcommand{\+}{{+\!\!\!+}}

\usepackage[top=2.75cm, bottom=2.75cm, left=2.75cm, right=2.75cm]{geometry}

\begin{document}
\thispagestyle{empty}

\begin{flushright} \small
IMPERIAL-TP-2008-CH-02
 \\UUITP-5/08 \\
 YITP-SB-08-21\\
\end{flushright}
\smallskip
\begin{center} \LARGE
{\bf    Euclidean Supersymmetry, Twisting and Topological Sigma Models}
 \\[5mm] \normalsize
{\bf C.M.~Hull $^{a,c}$, U.~Lindstr\"om$^{b}$, L.~Melo dos Santos$^{b,c}$,
R. von Unge$^{d,e,f}$,
and M.~Zabzine$^{b}$} \\[5mm]
 {\small\it
 $^a$The Institute for Mathematical Sciences\\
 Imperial College London\\
 53 Prince's Gate, London SW7 2PG, U.K.\\
 ~\\
$^b$Department of Theoretical Physics
Uppsala University, \\ Box 803, SE-751 08 Uppsala, Sweden \\
~\\
$^c$The Blackett Laboratory, Imperial College London\\
Prince Consort Road, London SW7 2AZ, U.K.\\
~\\
$^d$C.N.Yang Institute for Theoretical Physics, Stony Brook  
University, \\
Stony Brook, NY 11794-3840,USA\\
~\\
$^e$Simons Center for Geometry and Physics, Stony Brook University, \\
Stony Brook, NY 11794-3840,USA\\
~\\
$^{f}$Institute for Theoretical Physics, Masaryk University, \\
61137 Brno, Czech Republic \\~\\}
\end{center}
\vspace{2mm}
\centerline{\bfseries Abstract} \bigskip

\noindent We discuss two dimensional  $N$-extended supersymmetry in Euclidean signature and
its R-symmetry. For $N=2$, the R-symmetry is $SO(2)\times SO(1,1)$, so that only an $A$-twist is possible.
To formulate a $B$-twist, or to construct Euclidean $N=2$ models with $H$-flux so that the target geometry is generalised Kahler, it is necessary to work with a complexification of the sigma models.
These issues are  related to the obstructions to the existence of non-trivial twisted chiral superfields in Euclidean superspace.

\eject
\section{Introduction}
The construction of the topological sigma model by twisting the $(2,2)$
supersymmetric sigma model pioneered in  \cite{Witten:1988xj} and
further discussed in, e.g.,  \cite{Labastida:1991qq}, \cite{Clay},
explicitly or implicitly assumes the existence of an underlying
$(2,2)$ Euclidean supersymmetry. In this   letter we analyse such
supersymmetries and show that, strictly speaking, the $R$ symmetry
group does not allow for both an $A$ and a $B$ twist, but only an
$A$ twist. In 2D Lorentzian  space, (2,2) supersymmetry has
R-symmetry $SO(2)\times SO(2)$. One might expect that, after Wick
rotating so that the Lorentz group becomes $SO(2)$, the resulting
theory should have $SO(2)\times SO(2)\times SO(2)$ symmetry,
allowing one to twist the $SO(2)$ Lorentz symmetry with the diagonal
subgroup of  the $SO(2)\times SO(2)$ R-symmetry to give the $A$-twist
or the anti-diagonal subgroup of  the $SO(2)\times SO(2)$ R-symmetry
to give the $B$-twist. However, this Wick rotation  with  $SO(2)\times
SO(2)\times SO(2)$ symmetry gives a theory which is not
supersymmetric, so that the twisted versions would not automatically
have the desired BRST symmetry. Here we analyse (2,2) supersymmetry
in Euclidean 2D space, and find that the R-symmetry is not
$SO(2)\times SO(2)$ but  is instead $SO(2,\mathbb{C}) =SO(2)\times
SO(1,1)$. This allows an $A$-twist with the $SO(2)$ subgroup of the
R-symmetry, but not a $B$-twist.
 The $B$-twist requires going to the complexification of the theory. Indeed, this is implicit already in the early work on the subject. In 2D Euclidean space, left-handed and right-handed fermions are related by complex conjugation. The B-model has different twists for left and right-handed fermions, requiring them to be treated as independent so that one is formally dealing with the complexified model.

These issues can be, and usually are, suppressed in the discussion of topological sigma-models with
Calabi-Yau target spaces. However, they become important in discussing  topological sigma-models with H-flux, so that the target space has Generalized K\"ahler Geometry.
As in the analysis of the models in \cite{Hull:2008vs},    one needs to consider a complexified version of the Euclidean sigma model. One place where a careful treatment of these issues is particularly relevant is in understanding whether or not the Wess-Zumino term has the factor of \lq $i$' one would expect for a Euclidean sigma-model. Indeed, it was seen in  \cite{Hull:2008vs} that some parts of the
Wess-Zumino term in the twisted model are imaginary and have an interpretation in terms of gerbes, while others are real and contribute to a complexified Kahler class.
Another place where one can  see that there is a problem is in the (2,2) superspace formulation of the sigma-models.
In Lorentzian signature, the general  (2,2) sigma model can be written in terms of chiral, twisted chiral and semi-chiral superfields \cite{Lindstrom:2005zr}, \cite{Buscher:1987uw}.
In Euclidean signature, as we will review below, twisted chiral superfields are problematic
and there is no sensible way of continuing  twisted chiral superfields to Euclidean signature, unless
one goes to the complexified model.

\section{Supersymmetry Algebra}
The  $(p,q)$  Lorentzian (pseudo) supersymmetry algebra in $2D$ is given by\footnote{We have changed nomenclature from the original ``twisted-'' to ``pseudo-'' supersymmetry to avoid confusion when discussing another twist below.}\cite{Hull:1997kk},\cite{AbouZeid:1999em}
\beq
\{Q^I_+,Q^J_+\}&=&2i\eta^{IJ}\partial_\+~, I,J=1,...,p ~,\cr
\{Q^{I'}_-,Q^{J'}_-\}&=&2i\eta^{I'J'}\partial_=~,I',J'=1,...,q ~,\label{alg}
\eeq
where the supercharges $Q_\pm=Q^\dagger_\pm$ are  Majorana-Weyl spinors of   chirality $\pm 1$.
Ordinary supersymmetry corresponds to $\eta^{IJ}=\delta^{IJ}$, $\eta^{I'J'}= \delta ^{I'J'}$ while
for pseudo supersymmetry $\eta^{IJ},\eta^{I'J'}$ are
  arbitrary symmetric matrices, which we shall take to be invertible. (We shall not discuss the possibility of central charges here.)
The group of automorphisms of the algebra (\ref{alg}) include transformations
\beq
Q^I_+ &\to& M^I_{~J}Q^J_+\quad :M^t\eta M=\eta~,\cr
Q^{I'}_- &\to& \tilde M^{I'}_{~J'}Q^{J'}_-\quad :\tilde M^t\eta' \tilde M=\eta'~.\label{tfs}
\eeq
To preserve the Majorana-Weyl conditions, the matrices $M,\tilde M$ are real and independent.
Thus, the group of automorphisms is (space-time $\times ~ R$-symmetry)
\beq
SO(1,1)\times SO(n,p-n)\times SO(m,q-m)~,
\eeq
where $n (m)$ denotes the number of positive eigenvalues of $\eta ~ (\eta')$.
For ordinary $(p,q)$  supersymmetry with   $\eta=\delta$ and $\eta'=\delta$, the group  is
\beq
SO(1,1)\times SO(p)\times SO(q)~.
\eeq

In Euclidean signature there are no Majorana-Weyl fermions but we may use complex Weyl fermions. Hermitian conjugation changes the chirality according to
\beq
(Q_\pm)^\dagger= Q_\mp~.
\eeq
This means that we must have an equal number of left and right supersymmetries, $p=q:=N$. Since   the charges are now complex, the R-symmetry transformations can be generalised to allow the matrices $M$ in  (\ref{tfs}) to be complex. Then the R-symmetry transformations are
\beq
Q^I_+ &\to& M^I_{~J}Q^J_+\quad :M^t\eta M=\eta~.
 \label{tfse}
\eeq
This implies that the complex matrices $M\in SO(N,\mathbb{C})$, so that
 in Euclidean signature, the   group of automorphisms of $N$-extended supersymmetry is
\beq
SO(2)\times SO(N,\mathbb{C})~.
\eeq
Note that the negative chirality supercharges transform under the complex conjugate transformations
\beq
Q^{I}_- &\to& \bar M^{I}_{~J}Q^J_- ~.\label{tfsa}
\eeq

\section{Twisting and Sigma Models}

Twisting an $N=2$ supersymmetric  Euclidean theory in $2D$ involves
selecting an $SO(2)$ subgroup of the $R$ symmetry group
and then twisting the $2D$ Lorentz group $SO(2)$ with the $SO(2)$ R-symmetry subgroup, so that the new Lorentz group is an $SO(2)$ subgroup of this $SO(2)\times SO(2)$.
  We see from the above that the
$R$ symmetry group for the $(2,2)$ model is
\beq
SO(2,\mathbb{C})=SO(2)\times SO(1,1)~.\label{Rsym}
\eeq
There is then a unique choice of $SO(2)$ subgroup of the R-symmetry group, and twisting with this gives an $A$-twist.
A $B$-twist is not possible for realisations of this supersymmetry, as in going to the Euclidean theory, the second $SO(2)$ of the Lorentzian R-symmetry has become an $SO(1,1)$.

We now turn to the application of our discussion to (2,2)
supersymmetric sigma models. A useful starting point is the $N=1$
supersymmetric sigma-model in 4D Lorentzian spacetime. This has a
Kahler target space and $SO(2)$ R-symmetry \cite{Zumino}. It can be
formulated in terms of chiral superfields $\phi$, with $N=1$
superspace lagrangian given by the Kahler potential $K(\phi, \bar
\phi)$. Dimensionally reducing from $3+1$ dimensions on two
spacelike dimensions gives a theory in $1+1$ dimensions with
R-symmetry $SO(2)\times SO(2)$, with the extra $SO(2)$ arising from
rotation symmetry in the two internal dimensions. Alternatively,
reducing on one space and one time dimension gives a theory in two
Euclidean dimensions with R-symmetry $SO(2)\times SO(1,1)$, with the
extra $SO(1,1)$  arising from Lorentz transformations in the two
internal dimensions. In both cases, dimensional reduction ensures
$N=2$ supersymmetry in the reduced theory, and the reduction gives a
natural understanding of the R-symmetry groups in the two cases. In
both cases, the theory can be written in $N=2$ superspace in terms
of chiral superfields $\phi$ and their complex conjugates,
anti-chiral superfields $\bar \phi$ satisfying the constraints 
\beq\label{chir}
\bar {\mathbb{D}}_\pm \phi=0~,\quad \mathbb{D}_\pm\bar \phi =0~,
\label{chiral}
 \eeq 
 with the standard  supercovariant derivatives
\beq
\left\{\mathbb{D}_\pm , \bar{\mathbb{D}}_\pm\right\}=2i\partial_{\stackrel \+ =}~.
\eeq
In Euclidean signature $\partial_\+\to\partial, \partial_=\to\bar \partial$. The chirality constraints (\ref{chir}) make sense in Euclidean
signature as well as Lorentzian.

Consider now the extension of these models to include a Wess-Zumino
term. For the  $N=2$ sigma model on a Lorentzian 2D world-sheet, the
target space is then a bihermitian geometry \cite{Gates:1984nk},
recently recast as a generalised Kahler geometry \cite{gualtieri}.
The off-shell models of \cite{Gates:1984nk} are formulated in $N=2$
superspace with both chiral superfields $\phi$ and
 twisted chiral superfields $\chi$, which
satisfy the Lorentzian constraints
\beq\label{twist}
\bar {\mathbb{D}}_+ \chi=\mathbb{D}_-\chi=0~,\quad \mathbb{D}_+\bar \chi =\bar {\mathbb{D}}_-\bar \chi =0~.\label{twisted}
\eeq
The superspace lagrangian is then a generalised Kahler potential
$K(\phi, \bar \phi, \chi, \bar \chi )$.
 A complex coordiante transformation in superspace exchanges the constraints (\ref{chiral}) and (\ref{twisted}). A model in 1+1 dimensions with {\em only} twisted chiral fields is thus equal to one with {\em only} chiral fields.  The general case has semi-chiral superfields as well
as chiral   and
 twisted chiral superfields \cite{Lindstrom:2005zr}.

The natural expectation would be that the version of these models with Euclidean world-sheet should again have  chiral   and
 twisted chiral superfields. However, there is a problem with twisted chiral superfields in superspace, as was first realised in
 \cite{Buscher:1987uw}.
In Euclidean signature, the conjugation relations $(\mathbb{D}_\pm)^\dagger= \bar {\mathbb{D}}_\mp$ imply that conjugating the  constraints
(\ref{twisted})
give
\beq\label{twista}
\bar {\mathbb{D}}_- \chi=\mathbb{D}_+\chi=0~,\quad \mathbb{D}_-\bar \chi =\bar {\mathbb{D}}_+\bar \chi =0~, \label{twisteda}
\eeq
which together with (\ref{twisted})
 force $\chi, \bar \chi$ to be constant.
If instead one takes the constraints
 $\bar {\mathbb{D}}_+ \chi=\mathbb{D}_-\chi=0$ plus their conjugates $ \mathbb{D}_-\bar \chi =\bar {\mathbb{D}}_+\bar \chi =0$, then only the $\chi$-independent part of the potential $K(\phi, \bar \phi, \chi, \bar \chi )$ contributes to the geometry, and  this reduces to the usual Kahler case in terms of chiral superfields only.
There is one final possibility that does not involve complexifying  the twisted chiral fields.
That is to have a real superfield $ \chi$ satisfying the twisted chiral constriant
$\bar {\mathbb{D}}_+ \chi=\mathbb{D}_-\chi=0$ and an independent real twisted anti chiral superfield $\tilde \chi$ satisfying
$\mathbb{D}_+\tilde \chi =\bar {\mathbb{D}}_-\tilde \chi =0$. The superspace lagrangian $K(\phi, \bar \phi, \chi, \tilde \chi )$ then
 gives an interesting (2,2) sigma-model in Euclidean space, but with a target space of indefinite signature which is not generalised Kahler; this model will be discussed elsewhere \cite{HLMUZ}.
 
 So far we have limited the discussion to $(2,2)$ sigma models described by chiral and twisted chiral fields. To describe a general $(2,2)$ model, semi-chiral superfields are also needed \cite{Buscher:1987uw}, \cite{Lindstrom:2005zr}.  In Lorentzian signature  the left and right semi-(anti)chiral superfields obey the constraints
 \beq
 \bar {\mathbb{D}}_+ \mathbb{X}_L=0~,~~~ \bar {\mathbb{D}}_- \mathbb{X}_R = 0~,~~~
 \mathbb{D}_+ \bar{\mathbb{X}}_L=0~,~~~ \mathbb{D}_- \bar{\mathbb{X}}_R =0~.
 \label{semis}
 \eeq
 and 
 there is a local formulation in terms of chiral, twisted chiral and semi-chiral superfields, with
 a generalised Kahler potential
$K(\phi, \bar \phi, \chi, \bar \chi ,  \mathbb{X}_L,\bar  \mathbb{X}_L,   \mathbb{X}_R,\bar  \mathbb{X}_R)$. 
A potential depending on only one kind of semi-chiral superfield, 
$K(\phi, \bar \phi, \chi, \bar \chi ,  \mathbb{X}_L,\bar  \mathbb{X}_L)$ say, does not have a standard kinetic term for the components of $ \mathbb{X}_L$, so that the model has a topological nature in this sector \cite{Buscher:1987uw}.

  For world-sheets of Euclidean signature, one can similarly introduce semi-chiral fields $\mathbb{Y}_L, \mathbb{Y}_R$, but now the constraints consistent with complex conjugation are
  \beq
 \bar {\mathbb{D}}_+ \mathbb{Y}_L=0~,~~~ \bar {\mathbb{D}}_- \mathbb{Y}_R = 0~,~~~
 \mathbb{D}_+ \bar{\mathbb{Y}}_R=0~,~~~ \mathbb{D}_- \bar{\mathbb{Y}}_L =0~.
 \label{esemis}
 \eeq
 Now  a generalised Kahler potential
$K(\phi, \bar \phi, \chi, \bar \chi ,  \mathbb{Y}_L,\bar  \mathbb{Y}_L,   \mathbb{Y}_R,\bar  \mathbb{Y}_R)$ gives a kinetic term  for the components of  the semi-chiral superfields which is non-positive, constructed from a metric of indefinite signature. 
The change in the constraints  means that 
a potential depending on only one kind of semi-chiral superfield, 
$K(\phi, \bar \phi, \chi, \bar \chi ,  \mathbb{Y}_L,\bar  \mathbb{Y}_L)$ gives a  standard kinetic term for the components of $ \mathbb{Y}_L$.  
The geometry of these models containing semi-chiral fields will be discussed elsewhere.

\section{Complexification}

In order to formulate a Euclidean version of the supersymmetric sigma models with generalised Kahler targets, or to formulate a $B$-twist, it is necessary to work with   complexified theories in which
positive and negative chirality fields are treated as independent and are no longer complex conjugate, as they would be in Euclidean space.
The complex world-sheet coordinates $z,\bar z$ are treated as independent complex variables rather than as complex conjugates (as often done in conformal field theory), and the metric
\beq
ds^2= 2 dz d\bar z
\eeq
is preserved by the complexified Lorentz group
\beq
SO(2,\mathbb{C}) \simeq \mathbb{C}^* \simeq SO(2)\times SO(1,1)
\eeq
under which
$z\to  az$, $\bar z \to a^{-1} \bar z$ for $a\in \mathbb{C}^*$.
The positive chirality supercharges $Q_+$ are regarded as independent of the negative chirality ones
$Q_-$, so that again we can have $(p,q)$ supersymmetry with algebra (\ref{alg}).
The automorphisms are again of the form (\ref{tfs}) but with $M^I_{~J}$ and  $\tilde M^{I'}_{~J'}$ independent complex matrices, so that the R-symmetry group is
$SO(p,\mathbb{C})\times SO(q,\mathbb{C})$, and the full symmetry group is
\beq
SO(2,\mathbb{C}) \times SO(p,\mathbb{C})\times SO(q,\mathbb{C})~.
\eeq
In particular, for (2,2) supersymmetry, this group becomes
\beq
SO(2,\mathbb{C}) \times SO(2,\mathbb{C})\times SO(2,\mathbb{C})
\eeq
and allows both an $A$-twist and a $B$-twist, as well as a half-twist.

In superspace, one can introduce
chiral superfields $\phi$ and independent  anti-chiral superfields $\bar \phi$ satisfying the constraints
(\ref{chiral})
together with  twisted chiral superfields $\chi$  and independent twisted 
 anti-chiral superfields $\bar \chi $ satisfying the constraints
(\ref{twisted})
and the   superspace lagrangian is again a generalised Kahler potential
$K(\phi, \bar \phi, \chi, \bar \chi )$. This is consistent so long as $\phi, \bar \phi, \chi, \bar \chi$ are treated as independent complex fields, and gives a target geometry which is a complexification of generalised Kahler geometry.
This allows both an $A$-twist and a $B$-twist, and it was the twisting of this complexified sigma-model that was analysed in \cite{Hull:2008vs}.

Similarly, one can introduce  left and right semi-chiral superfields 
$ \mathbb{X}_L,   \mathbb{X}_R$ and independent  anti-semi-chiral ones $ \bar  \mathbb{X}_L,\bar  \mathbb{X}_R$ satisfying the constraints (\ref{semis}).
Then the general  superspace lagrangian is given by a generalised Kahler potential
$K(\phi, \bar \phi, \chi, \bar \chi ,  \mathbb{X}_L,\bar  \mathbb{X}_L,   \mathbb{X}_R,\bar  \mathbb{X}_R)$. 
This complexified geometry gives a Euclideanisation of the standard Lorentzian signature sigma model with generalized K\"ahler target geometry.
 From the superspace
point of view,   when all fields are complexified we can have
chiral, twisted chiral and semi-chiral superfields in the model. This is
necessary, e.g., to be able to discuss twisting, mirror symmetry or T-duality
in superspace. We plan to return to Euclidean $(2,2)$ sigma models
in superspace  in a separate publication \cite{HLMUZ}
\bigskip

{\bf Acknowledgement:} UL acknowledges support by EU grant
(Superstring theory) MRTN-2004-512194 and by VR grant 621-2006-3365.The research of R.v.U. was supported by Czech ministry of education contract No. MSM0021622409. The research of M.Z. was supported 
by VR-grant 621-2004-3177. 
L.M.d.S. acknowledges support by FCT grant SFRH/BD/10877/2002.

\eject

\end{document}